%% file: main.tex
\documentclass[sigconf]{acmart}
\renewcommand\footnotetextcopyrightpermission[1]{}
\usepackage{booktabs} 
\usepackage[justification=centering]{caption}

\begin{document}
\title{Identifying Retweetable Tweets with a Personalized Global Classifier}

\author{Michail Vougioukas}
\affiliation{%
  \institution{blue-infinity S.A.}
  \city{Geneva} 
  \state{Switzerland} 
}

\author{Ion Androutsopoulos}
\affiliation{%
  \institution{Athens University of Economics and Business}
  \city{Athens} 
  \state{Greece} 
}

\author{Georgios Paliouras}

\affiliation{%
  \institution{National Center of Scientific Research ``Demokritos''}
  \city{Athens}
  \state{Greece}
}

\renewcommand{\shortauthors}{M. Vougioukas et al.}

\begin{abstract}
In this paper we present a method to identify tweets that a user may find interesting enough to retweet. The method is based on a global, but personalized classifier, which is trained on data from several users, represented in terms of user-specific features. Thus, the method is trained on a sufficient volume of data, while also being able to make personalized decisions, i.e., the same post received by two different users may lead to different classification decisions. Experimenting with a collection of approx.\ 130K tweets received by 122 journalists, we train a logistic regression classifier, using a wide variety of features: the content of each tweet, its novelty, its text similarity to tweets previously posted or retweeted by the recipient or sender of the tweet, the network influence of the author and sender, and their past interactions. Our system obtains $F_1 \approx 0.9$ using only 10 features and 5K training instances.
\end{abstract}

\maketitle
\input{body}

\bibliographystyle{ACM-Reference-Format}
\bibliography{refs} 
\end{document}

%% file: body.tex
\section{Introduction}

Information shared on social networks is ever increasing and users are often overwhelmed by the number of posts (e.g., tweets) they receive. Many of the incoming posts are of marginal or no interest to their recipients. Consequently, interesting posts may be ignored or overlooked by time-constrained users, who may also give up reading their timelines. Filters that estimate the interest of each incoming post can alleviate this problem, for example by allowing users to sort incoming posts by predicted interest (e.g., `top stories' vs.\ `most recent' in Facebook) or by mixing recent posts with predicted interesting ones (e.g., `in case you missed it' in Twitter).

There have been two main approaches to detect interesting posts in social networks: \emph{global} filters \cite{alonso_o_1,alonso_o_2,yang_m} and \emph{personal} filters \cite{waldner_w,vougioukas_m,chen_j}. Global filters try to predict how interesting a post is for the entire social network or at least a broad audience. A single global filter is typically trained on a large collection of posts and the reactions of all users to each post (e.g., total number of retweets per post). The trained global filter is then used to assign a single, user-independent interest score to each new post. By contrast, personal filters are typically trained on posts received by a particular user and the reactions of the particular user (e.g., whether or not the user retweeted each post). A separate filter is trained per user and is then employed to provide user-specific interest scores for each tweet or, generally, social post. Personal filters can, at least in principle, provide recommendations tailored to a particular user's own interests, which may not coincide with the interests of the majority of users that global filters are trained to predict. On the other hand, global filters are typically trained on much larger datasets compared to personal filters. Hence, global filters may work better in practice, especially with new users, for which personal filters may have very few training instances (the `cold start' problem).

\begin{figure}
\includegraphics[scale=0.4]{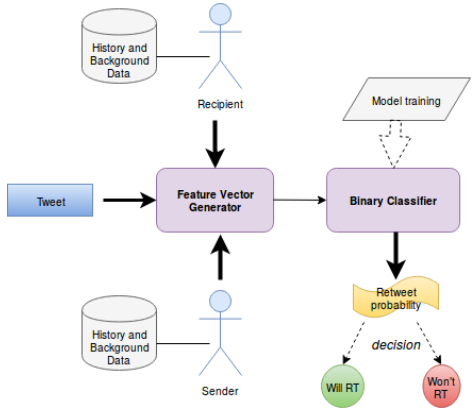}
\caption{Architecture of our system.}
\label{fig:system}
\end{figure}

Following Uysal and Croft \cite{uysal_i} and Zhang et al.\ \cite{zhang_q}, in this paper we investigate a hybrid approach that attempts to combine the strengths of both global and personal filters. As in global filters, we train a \emph{single} system on a large collection of tweets received by multiple users. Each tweet, however, is represented as a feature vector that includes \emph{user-specific features} (Fig.~\ref{fig:system}), for example indicating the extent to which the incoming tweet is similar to tweets previously posted or retweeted by the recipient, or how often the recipient has retweeted posts of the sender of the tweet. If the same tweet is received by two different users, it will be represented by two different feature vectors. This allows the system to take into account user preferences and produce different predictions per recipient, even for the same incoming tweet, as in personal filters, while still being able to generalize over different users (e.g., learn that users are in general more likely to retweet posts that are similar to their own posts). We train a single shared logistic regression model for all users, in order to predict if a tweet received by a particular user will be retweeted by that user or not. We examine the effect of several types of features that examine the content of each incoming tweet, the similarity of the incoming tweet to tweets previously posted or retweeted by the recipient or the sender, the network influence of the sender and recipient, the interaction between them (e.g., if they have mentioned each other in previous tweets), the novelty of the incoming tweet (e.g., its similarity to tweets recently seen by the recipient). On a dataset of approx.\ 130K tweets received by 122 journalists, our system obtains $F_1 \approx 0.9$ using only 10 features and approximately 5K training instances. 

Using previous retweet (and non-retweet) actions as gold labels has the advantage that no extra human labeling is required to construct training and test data, as opposed to asking users to label their incoming tweets with interest scores. On the other hand, retweeting is only an approximate signal of interest, as users do not retweet all the posts they find interesting. Nevertheless, retweeting is usually an indication of great interest in a  post and, hence, our system can be used to detect tweets that a particular user would find very interesting (interesting enough to retweet), which could then be ranked higher or mixed with recent tweets. 

The main contributions of this paper are: (a) a lightweight prediction model, which attains high F1 score with a small number of features and training instances; (b) investigation of most candidate features mentioned in related literature and variants thereof, grouped into feature types for further research; (c) a large dataset of tweets and associated user information, which we plan to make publicly available in an encoded form.\footnote{Instructions to obtain the dataset will be made available at \url{http://nlp.cs.aueb.gr/}.}

Section~\ref{sec:systemDescription} below describes our system. Section~\ref{sec:experiments} presents the experiments we performed. Section~\ref{sec:related} discusses related work. Section~\ref{sec:conclusions} concludes and proposes future work. A summary of the work of this paper has also been published \cite{Vougioukas2017}.

\section{System description} \label{sec:systemDescription}

\subsection{System overview}

Our system predicts how likely it is that a particular user (the \emph{recipient} of Fig.~\ref{fig:system}) will retweet a particular incoming tweet. The system also has access to the history of the recipient (e.g., tweets the recipient has previously received or posted), the history of the sender of the tweet, as well as background information about the recipient and the sender (e.g., number of followers).\footnote{We use Twitter's API (\url{https://dev.twitter.com/rest/public}) to obtain this information.} By \emph{sender} we mean the user that caused the recipient to receive the tweet, either by authoring it directly (if the recipient follows the author) or by retweeting it (if the recipient does not follow the author). The tweet is represented as a feature vector, which includes features that depend on the particular recipient; hence, the same tweet will be represented by a different feature vector when the system tries to estimate if another recipient will retweet it or not. The feature vector is passed on to a (binary) logistic regression classifier that predicts if the recipient will retweet the incoming tweet or not. The classifier (one model for all recipients) is trained on tweets received by Twitter users and the users' reactions (whether they retweeted the incoming tweets or not).\footnote{We used Weka's implementation of logistic regression (\url{http://www.cs.waikato.ac.nz/ml/weka/}), with default hyper-parameter values. Modifying the defaults had no significant effect in preliminary experiments.}

\subsection{Preprocessing of the tweet text}

Before further processing, the text of each tweet is normalized as follows to allow the classifier to generalize (e.g., over different URLs, different numbers, smileys that express the same sentiment).

\begin{enumerate}
\small
\item All URLs are replaced by the same pseudo-token (e.g., `\texttt{\_url\_}'), which denotes a generic URL.
\item All numbers are replaced by a pseudo-token (e.g., `\texttt{\_num\_}'). 
\item Each type of smiley is replaced by a different pseudo-token: 
\begin{enumerate}
\small
\item Love/like smileys (e.g., `\texttt{<3}').
\item Positive sentiment smileys (e.g., `\texttt{:-)}').
\item Negative sentiment smileys (e.g., `\texttt{:-(}').
\item Neutral sentiment smileys (e.g., `\texttt{:-|}'). 
\end{enumerate}
\item All tokens are converted to lower case.
\end{enumerate}

These steps are based on the preprocessing used in GloVe \cite{pennington_j} to turn words into embeddings \cite{mikolov_t}. Hence, in a future extension of our system one could easily use GloVe embeddings.

\subsection{Features used by the classifier} \label{subsection:datarep_feat}

The feature vector of each incoming tweet contains up to 50 features, each corresponding to a factor that we suspect may help predict if the tweet will be retweeted or not. The features were constructed by taking into account previous related work (Section~\ref{sec:related}), the information provided by Twitter's API, and our own experience as Twitter users. The 50 features are divided into 7 groups. 

\begin{figure}
\includegraphics[scale=0.4]{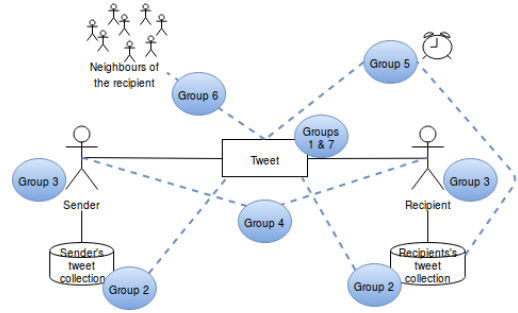}
\caption{Groups of features used by our system and how they relate to the tweet itself, the sender, the recipient etc.}
\label{fig:featurespace}
\end{figure}

Group 1 (Fig.~\ref{fig:featurespace}, Table~\ref{tab:feat1}) contains features that examine the tweet itself (e.g., length, if it contains a URL or not, if it mentions a Twitter account). Longer tweets, or tweets that contain URLs of longer posts (e.g., news articles) or photographs may be more informative and, thus, more interesting. Tweets that mention other user accounts may be parts of dialogues, which may be uninteresting to recipients, unless they interact frequently with the sender (see also Group 4). Hashtags may indicate trending topics. Tweets that have already been retweeted or favoured by many users are more likely to be important. Exclamation marks indicate surprise or strong feelings.

Group 2 (Fig.~\ref{fig:featurespace}, Table~\ref{tab:feat2}) contains features that examine how similar the incoming tweet is to particular collections of tweets (e.g., all tweets previously posted by the sender). The similarity between the incoming tweet $t$ and a collection of tweets $C = \{c_1, \dots, c_n\}$ is computed as the average TF-IDF cosine similarity between $t$ and each  $c_i$. The intuition in Group 2 is that recipients may prefer tweets that are similar or dissimilar (if they prefer surprising posts) to the posts of the particular sender, or their own posts, or the posts they usually see or retweet. 

Group 3 (Fig.~\ref{fig:featurespace}, Table~\ref{tab:feat3}) contains features modeling the network influence, popularity, and authority of the sender and the recipient. These features include Twitter account statistics (number of followers, number of posts, days active for, list subscriptions), features that may indicate authority (verified accounts, URLs in the description fields of their profiles), as well as scores obtained from Klout, a service that estimates a user's social influence by taking into account their activity in various social networks.\footnote{See \url{http://klout.com/}. All the features are normalized to $[0, 1]$.}

Group 4 (Fig.~\ref{fig:featurespace}, Table~\ref{tab:feat4}) contains features that capture the interaction between the sender and the recipient (e.g., whether or not tweets of the sender mention the recipient). The intuition is that recipients are more likely to be interested in posts of senders they interact more closely with. 

Group 5 (Fig.~\ref{fig:featurespace}, Table~\ref{tab:feat5}) contains features that attempt to estimate the timeliness of the incoming tweet. A tweet that is very similar to other recently received or retweeted tweets may be old news. The similarity scores of these features are again averaged TF-IDF cosine similarities.

Group 6 (Fig.~\ref{fig:featurespace}, Table~\ref{tab:feat6}) contains features related to the users the recipient follows (the user's {\em neighbours}). The neighbours presumably have common interests with the recipient. Hence, if the original author of the incoming tweet is a neighbour of the recipient or if the incoming tweet has been retweeted by many neighbours of the recipient, this may be an indication that the recipient will also find the incoming tweet interesting.

Group 7 (Fig.~\ref{fig:featurespace}, Table~\ref{tab:feat7}) complements the features of Group 1 by looking for particular keywords and parts of speech (nouns, verbs, articles) in the incoming tweet.\footnote{We use CMU ARK Twitter tagger \cite{gimpel_k} (\url{http://www.cs.cmu.edu/~ark/TweetNLP/}).} The features of Group 7 are based on the work of Tan et al. \cite{tan_c}, who found that the wording of a post significantly affects its propagation, compared to other posts that express the same information using different wordings. Tan et al.\ provide a list of 20 `good' keywords, believed to increase the propagation probability of a post, and 20 `bad' keywords.

\begin{table}
  \caption{Features of Group 1 (the tweet itself).}
  \label{tab:feat1}
  \begin{tabular}{{p{0.17\linewidth}p{0.77\linewidth}}}
    \hline
    Feature ID & Feature Description\\\hline
     FT1 & Tweet length in characters.\\
     FT2 & Does the tweet contain a URL?\\
     FT3 & Does it mention a Twitter account ($@$username)?\\
     FT4 & Does it contain a hashtag?\\
     FT5 & Global retweet count (times it has been retweeted). \\
     FT6 & Global favourite count.\\
     FT7 & Does the tweet contain an exclamation mark?\\
     FT8 & Does it contain a photo?\\
     FT9 & Number of Twitter accounts it mentions.\\\hline
\end{tabular}
\end{table}

\begin{table}
  \caption{Features of Group 2 (average TF-IDF cosine similarity of the tweet to other tweet collections).}
  \label{tab:feat2}
  \begin{tabular}{{p{0.16\linewidth}p{0.78\linewidth}}}
    \hline
    Feature ID & Feature Description\\\hline
     FT10 & Similarity to tweets previously posted (authored or retweeted) by the sender.\\
     FT11 & Similarity to tweets previously posted (authored or retweeted) by the recipient.\\
     FT12 & Similarity to tweets previously seen by the recipient (excluding `easy' negative tweets and tweets from recently inactive neighbours -- see Section~\ref{sec:dataset}). \\
     FT13 & Similarity to previous retweets of the recipient.\\\hline
\end{tabular}
\end{table}

\begin{table}
  \caption{Features of Group 3 (influence, popularity, authority of the sender and recipient).}
  \label{tab:feat3}
  \begin{tabular}{{p{0.16\linewidth}p{0.78\linewidth}}}
    \hline
    Feature ID & Feature Description\\\hline
     FT14 & Number of users that follow the sender.\\
     FT15 & Number of users the sender follows.\\
     FT16 & Number of tweets the sender has posted (authored or retweeted).\\ 
     FT17 & Number of curated lists the sender subscribes to.\\
     FT18 & Is the sender a verified account?\\
     FT19 & Days the sender's account has been active for.\\
     FT20 & Does the sender have a URL in their description?\\
     FT21 & The Klout score (influence) of the sender.\\
     FT22 & Delta of FT21 from the previous 24 hours.\\
     FT23 & Delta of FT21 from the previous 7 days.\\
     FT24 & Delta of FT21 from the previous 30 days.\\
     FT25 & Number of users that follow the recipient.\\
     FT26 & Number of users the recipient follows.\\
     FT27 & Number of tweets the recipient has posted.\\
     FT28 & Number of curated lists the recipient subscribes to.\\
     FT29 & Is the recipient a verified account?\\
     FT30 & Days the recipient's account has been active for.\\
     FT31 & Does the recipient have a URL in their description?\\
     FT32 & The Klout score of the recipient.\\
     FT33 & Delta of FT32 from the previous 24 hours.\\
     FT34 & Delta of FT32 from the previous 7 days.\\
     FT35 & Delta of FT32 from the previous 30 days.\\\hline
\end{tabular}
\end{table}

\begin{table}
  \caption{Features of Group 4 (sender-recipient interaction).}
  \label{tab:feat4}
  \begin{tabular}{{p{0.16\linewidth}p{0.78\linewidth}}}
    \hline
    Feature ID & Feature Description\\\hline
     FT36 & Is the recipient mentioned ($@$username) in the incoming tweet?\\
     FT37 & Has the sender ever mentioned the recipient?\\
     FT38 & Has the recipient ever mentioned the sender?\\
     FT39 & Has the sender ever retweeted the recipient?\\
     FT40 & Has the recipient ever retweeted the sender?\\
     FT41 & No.\ of times the recipient has retweeted the sender.\\\hline
\end{tabular}
\end{table}

\begin{table}
  \caption{Features of Group 5 (timeliness of incoming tweet).}
  \label{tab:feat5}
  \begin{tabular}{{p{0.16\linewidth}p{0.78\linewidth}}}
    \hline
    Feature ID & Feature Description\\\hline
     FT42 & Similarity to tweets seen by the recipient during the previous week  (excluding `easy' negative tweets and tweets from recently inactive neighbours). \\
     FT43 & Similarity to tweets retweeted by the recipient during the previous week.\\\hline
\end{tabular}
\end{table}

\begin{table}
  \caption{Features of Group 6 (neighbours of the recipient).}
  \label{tab:feat6}
  \begin{tabular}{{p{0.16\linewidth}p{0.78\linewidth}}}
    \hline
    Feature ID & Feature description\\\hline
     FT44 & Is the author of the incoming tweet a neighbour of the recipient? (The sender may be the author of the tweet or a neighbour that retweeted it. In the latter case, the original author may not be a neighbour.) \\
     FT45 & Number of times the incoming tweet has been retweeted by the neighbours of the recipient.\\\hline
\end{tabular}
\end{table}

\begin{table}
  \caption{Features of Group 7 (wording of the tweet).}
  \label{tab:feat7}
  \begin{tabular}{{p{0.16\linewidth}p{0.78\linewidth}}}
    \hline
    Feature ID & Feature Description\\\hline
     FT46 & Number of keywords in the incoming tweet explicitly asking to retweet/share (e.g., `RT', `spread', `share').\\
     FT47 & Number of nouns and verbs in the incoming tweet. \\
     FT48 & Number of definite articles in the incoming tweet. \\
     FT49 & Number of indefinite articles in the incoming tweet.\\
     FT50 & Number of `good' keywords minus number of `bad' keywords in the tweet, using the keywords of \cite{tan_c}. \\\hline
\end{tabular}
\end{table}

\section{Experiments} \label{sec:experiments}

\subsection{Dataset} \label{sec:dataset}

In our experiments, the recipients (Fig.~\ref{fig:system} and \ref{fig:featurespace}) were 122 journalists. We started with a list of 262 journalists, available from previous work \cite{zamani_k}, but we retained only  journalists that write in English.\footnote{We used a flag in Twitter's API to detect the language.} We also discarded journalists for which we could not collect at least 500 retweets, ending up with 122 journalists. The dataset of our experiments consists of 122 subsets, one for each journalist. Each subset comprises the most recent retweets of the corresponding journalist that we could collect through Twitter's API. The number of retweets in each subset was at least 500 and at most 2,500.\footnote{We could not collect more, due to restrictions of Twitter's API.} In each subset, the journalist's retweets are treated as \emph{positive instances}. 

Each subset also contains \emph{negative instances}, meaning incoming tweets that the journalist did not retweet. To obtain the negative instances for each journalist we crawled the timelines of the users the journalist follows (neighbours) and collected their most recent posts (tweets authored or retweeted by the neighbour) that were not included in the positive instances of the journalist. To make the dataset more challenging, we excluded \emph{`easy' negative instances}, meaning incoming tweets from neighbours that the journalist has never retweeted in the past, assuming that the journalist does not really care about posts from such neighbours. We also excluded negative instances from \emph{recently inactive neighbours} (neighbours without any posts in the last seven days). 

Our dataset was collected in late September 2015. To avoid using very old tweets, we discarded instances that were posted before January 2014.  Hence, the dataset covers a period of approximately 19 months and contains approximately 12 million instances in total, involving 63,800 users (senders or recipients). Since the collected negative instances were many more than the positive ones, we randomly downsampled the negative instances of each journalist to obtain an equal number of positive and negative instances in each subset. This left a total of 133,000 instances (66,500 positive, 66,500 negative) in the 122 subsets.\footnote{IDF scores were estimated on the 12 million instances.} To create training, development, and test sets, we first merged the 122 subsets and temporally ordered (by time posted) all the positive instances and, separately, all the negative instances. We removed all incoming duplicates per receiver (e.g., same tweet reaching the same receiver at different times via retweets of different senders the receiver follows), keeping only the earliest among duplicates.

We then formed 140 temporally ordered \emph{batches}. Batch 1 contains the earliest 475 positive and the earliest 475 negative of the 133,000 instances. Batch 2 contains the next 475 positive and the next 475 negative instances etc.\footnote{The incoming tweets of the 122 journalists are distributed almost uniformly across the batches.} The first 120 batches were used as the \emph{training set} (57,000 positive and 57,000 negative instances), the next 10 batches were used as the \emph{balanced development set} (4,750 positive and 4,750 negative instances), and the last 10 batches were used as the \emph{balanced test set} (4,750 positive and 4,750 negative instances). We also constructed alternative, \emph{unbalanced development and test sets} by randomly downsampling the positive (retweeted) instances in each batch of the balanced development and test sets, leaving 25 positive (5\%) and 475 negative instances (95\%) in each batch (250 positive and 4,750 negative instances in each unbalanced set). 

We always train the logistic regression classifier of our system (Fig.~\ref{fig:system}) on the balanced training set. Using a balanced training set is common practice for discriminative supervised learning algorithms. Previous experiments \cite{vougioukas_m2} also indicated that training the logistic regression classifier on a balanced set leads to better performance on the development set, compared to using an unbalanced training set, even when the classifier is evaluated on an unbalanced development set with the same positive-to-negative ratio as the unbalanced training set. For a classifier trained on a balanced set, the balanced development and test sets are expected to be easier than their unbalanced counter-parts, since all the balanced sets have the same priors; this is also confirmed by our experimental results. The balanced development and test sets, however, are unrealistic, because they assume that receivers retweet on average half of their incoming tweets. The unbalanced development and test sets are intended to evaluate our system in a  more realistic scenario, where receivers retweet only 5\% of their incoming tweets. 

To bypass privacy issues, the training, development, and test sets (balanced and unbalanced) of our experiments will be made publicly available in an encoded form, where words will be replaced by unique integer identifiers, as in previous spam filtering and legal text analytics datasets we have made available \cite{androutsopoulos_i,Chalkidis2017}. We also plan to provide pre-trained word embeddings (e.g., generated by word2vec \cite{mikolov_t} or GloVe \cite{pennington_j}) for each encoded word (integer identifier).

\subsection{Incremental training and evaluation} \label{sec:incremental}

To study the effect of the size of the training set, each experiment was repeated 120 times, each time training the logistic regression classifier on the first (earliest) $k$ batches of the training set ($1 \leq k \leq 120$), always using the same development or test set (10 batches each) to evaluate the performance of the classifier for each $k$ value. We used \emph{precision} ($P$), \emph{recall} ($R$), and \emph{F1 score} to evaluate the performance of the classifier, defined as usually.

\subsection{Experiments on the development set} \label{sec:devExperiments}

To get a first view of the usefulness of the features of Section~\ref{subsection:datarep_feat}, we ranked them by  decreasing Pearson correlation \cite{benesty_j} to the class label, using a 10-fold cross-validation on the training set (Section~\ref{sec:dataset}). The Pearson correlations of the top 10 features are shown in Table~\ref{tab:correl}. Interestingly, the seven feature groups of Section~\ref{subsection:datarep_feat} are not equally represented in the top 10 (Table~\ref{tab:correl}). Only Group 2 (content similarity), Group 3 (influence, authority, popularity, but mostly of the sender), Group 4 (sender-recipient interaction), and Group 6 (neighbours) have features among the top 10. 

\begin{table}
  \caption{Pearson correlation of the top 10 features to the class label (10-fold cross-validation on the training set).}
  \label{tab:correl}
  \small
  \begin{tabular}{{p{0.11\linewidth}p{0.08\linewidth}p{0.64\linewidth}}}
    \hline
    Feature & Pearson & Feature Description\\\hline
     FT43 & 0.60 & Similarity to tweets retweeted by the recipient during the previous week.\\
     FT10 & 0.57 & Similarity to tweets previously posted (authored or retweeted) by the sender.\\
     FT21 & 0.49 & The Klout score (influence) of the sender.\\
     FT16 & 0.47 & Number of tweets the sender has posted.\\
     FT13 & 0.44 & Similarity to tweets previously retweeted by the recipient.\\
     FT45 & 0.42 & Number of times the tweet has been retweeted by the recipient's neighbours.\\
     FT44 & 0.40 & Is the author of the incoming tweet a neighbour of the recipient?\\
     FT40 & 0.40 & Recipient ever retweeted the sender?\\
     FT11 & 0.40 & Similarity to tweets previously posted (authored or retweeted) by the recipient.\\
     FT38 & 0.36 & Recipient ever mentioned the sender?\\\hline
\end{tabular}
\end{table}

\begin{figure}
\includegraphics[scale=0.35]{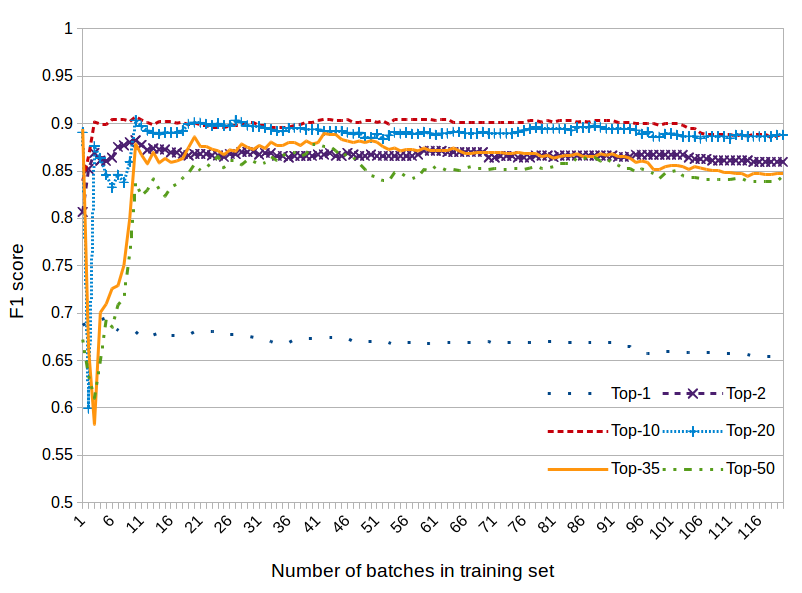}
\caption{F1 on the unbalanced development set, for different numbers of top features.}
\label{fig:exp1}
\end{figure}

We then evaluated the system with respect to its F1 score on the unbalanced development set, using an increasing number $k$ of training batches ($1 \leq k \leq 120$), with different numbers of top-$m$ features ($m \in \{1, 2, 10, 20, 35, 50\}$). The results of these experiments are shown in Fig.~\ref{fig:exp1}. A first observation is that the learning curves are steep for the first few training batches, but flatten out after approximately the first 12  batches (11,400 examples). This is a general trend for all of our experiments and suggests that a larger training set would not improve the system's performance.

A second observation is that the best results are obtained with the top 10 features (Fig.~\ref{fig:exp1}). Adding more features leads to increasingly worse results, possibly because the additional features add noise. Indeed, after the first 15-20 top features, the Pearson correlation of the features to the class label is quite low (\textless 0.13). The performance of a `lightweight' system with only the top two features (F1 $\approx$ 0.87) is comparable to that of the top 10 features (Fig.~\ref{fig:exp1}). 

\begin{figure}
\includegraphics[scale=0.35]{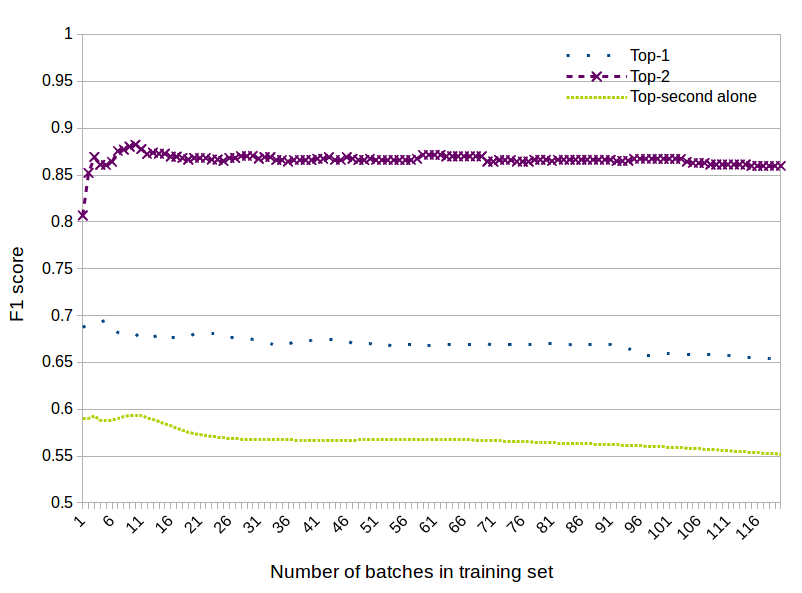}
\caption{F1 on the unbalanced development set, using only the top feature (FT10), only the 2nd-top (FT43), or both.}
\label{fig:exp2}
\end{figure}

We investigated further the notable change in F1 when the second top feature is added to the top one (Fig.~\ref{fig:exp1}, curves Top-1 and Top-2). Figure~\ref{fig:exp2} shows the F1 score, again on the unbalanced development set, using only the top feature (FT10), only the second-top (FT43), or both. The second-top feature alone is not a good predictor, but the combination of the two features increases F1.

\begin{figure}
\includegraphics[scale=0.35]{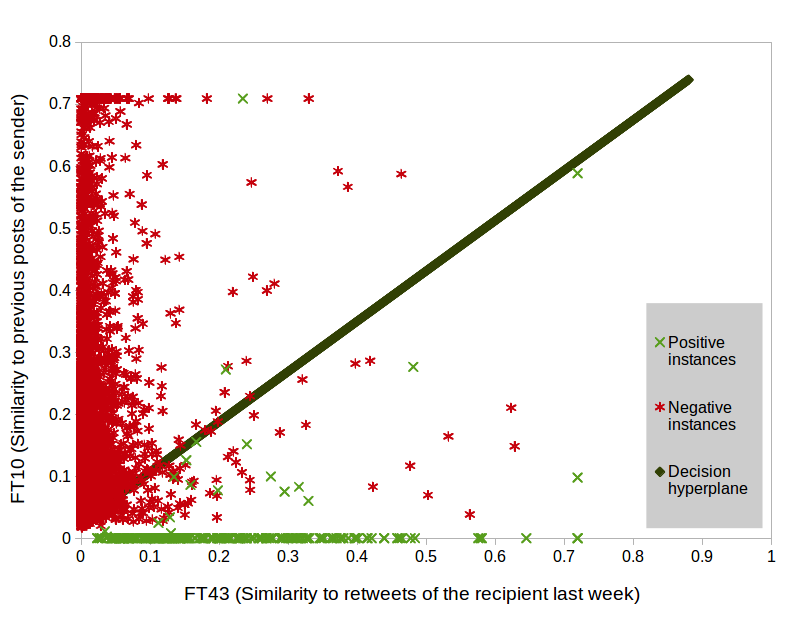}
\caption{Sample positive and negative instances from the unbalanced development set and the linear separator the logistic regression classifier learned on the training set.}
\label{fig:exp3}
\end{figure}

Figure~\ref{fig:exp3} sheds more light on the role of the top two features (FT10, FT43). It plots the positive and negative instances of a random subset (251 positive instances, 4,494 negatives) of the unbalanced development set. The straight line is the separator the logistic regression learned on the training set. In most cases, the line correctly separates the negative (stars) from the positive (crosses) instances, which agrees with the high F1 score in Figures~\ref{fig:exp1} and \ref{fig:exp2}. 

As one might expect, most negative instances (stars) have low similarity (small values on the horizontal axis of Fig.~\ref{fig:exp3}) to the tweets the recipient retweeted during the previous week (FT43). This suggests that recent retweets of the recipients are good indicators of their current interests. Perhaps more unexpectedly, most positive examples (crosses) have very \emph{low} similarities to the previous posts of the sender (FT10). Intuitively, recipients tend to prefer (or at least retweet) posts that are \emph{unusual} for the particular sender (posts that are surprisingly not about the usual topics of the sender, to the extent that TF-IDF cosine similarity captures topic similarity). 

Figure~\ref{fig:exp3} also illustrates the effect of combining the two features. Negative instances tend to have small values on the horizontal axis (FT43), but a non-negligible subset of positive instances also have small FT43 values. Most of those positive instances, however, have near-zero values on the vertical axis (FT10), unlike most negative instances and, hence, the combination of the two features improves classification accuracy. However, a non-linear classifier might manage to separate better the instances near the origin, where an S-shaped separator seems to be needed.

\subsection{Experiments on the test set} \label{sec:testExperiments}

\begin{figure}
\includegraphics[scale=0.35]{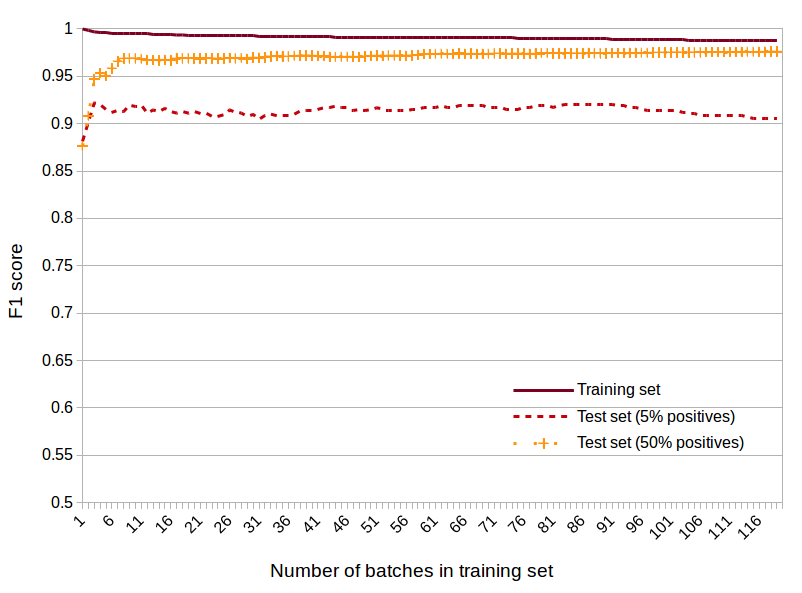}
\caption{F1 on the balanced and unbalanced test set vs.\ F1 on the (always balanced) training set, using the top 10 features.}
\label{fig:exp4}
\end{figure}

In a final set of experiments, we evaluated our system on the (previously unseen) test set (10 fresh batches), using both the balanced (50\% positives, 50\% negatives) and the unbalanced (5\% positives, 95\% negatives) versions of the test set (Section~\ref{sec:dataset}). We used the top 10 features in these experiments, which had led to the best results on the development set (Section~\ref{sec:devExperiments}). The training set was the same as in the previous experiments (balanced). Fig.~\ref{fig:exp4} shows the F1 scores on the two versions of the test set, along with the F1 scores on the batches of the training set the classifier has been trained on. The performance of a supervised classifier is typically better on the training data it has encountered, compared to its performance on unseen test data. Hence, the performance on the encountered training data is a boundary of the performance on test data. A large gap between the two is often due to overfitting the training data. The performance on the training data typically deteriorates as more training data are added, due to reduced overfitting.

Figure~\ref{fig:exp4} shows the system performs better on the unbalanced test set (F1 $\approx$ 0.92) than on the unbalanced development set (cf.\ Fig~\ref{fig:exp1}). As expected, the system performs better on the balanced test set, which has the same positive-to-negative ratio (50\% positives) as the training set, and worse on the unbalanced test set (5\% positives). The gap between the performance on the training and balanced test data is small, indicating that the system does not significantly overfit the training data. The larger gap between the performance on the training and unbalanced test data is due to the change of ratio from the training to the test data, which makes the problem more difficult for the classifier. Again, both test curves flatten out after very few training batches ($\sim$5 for the unbalanced test set, $\sim$12 for the balanced), though the balanced test F1 score continues to improve slowly, whereas the unbalanced test F1 does not.

\section{Related Work} \label{sec:related}

\subsection{Global filters for social media}

Global filters aim to identify content which is interesting for a large audience. Yang et al.\ \cite{yang_m} used Latent Dirichlet Allocation (LDA) in a filter aiming to detect globally interesting tweets, as opposed to tweets that are only interesting to their direct recipients. 

Hurlock and Wilson \cite{hurlock_j} investigated qualitative factors (e.g., reporting personal experience or not, providing specific information, timeliness, trusted author) that affect the perceived usefulness of the tweets returned by a search engine. Although they considered a different task (search) than the one we considered (predicting retweets) and their factors are not always easy to map to computable features (e.g., reporting personal experience, usefulness of a link), their work influenced our choice of features. 

Duan et al.\ \cite{duan_y} used a learning-to-rank algorithm, experimenting with several types of features. They found that features related to the authority of senders (e.g., number of lists the author is included in) along with tweet length and presence of URL were particularly useful. These findings influenced our choice of features. 

Alonso et al.\ \cite{alonso_o_1,alonso_o_2} considered several types of features and in their early work reported that a single feature (presence of URL) was enough to obtain 80\% accuracy. Their later work \cite{alonso_o_1}, however, showed that human annotators did not agree on which tweets were interesting (inter-annotator agreement was as low as for random choices), concluding that interest is a subjective, not global notion.

\subsection{Personal filters for social media}

In previous work \cite{vougioukas_m}, we developed personal filters for Twitter, using the incoming tweets of six recipients,  annotated with interest scores by the recipients themselves. Each filter was trained and tested on incoming tweets of a  particular recipient, using the same learning algorithm and features. Manual annotation turned out to be a bottleneck and we could not obtain more than 1,000 annotated incoming tweets per recipient. Thus, we concluded that training a separate filter per user is not realistic and does not address the cold start problem, where a filter must be provided to a new user (recipient), with no training data available for this user.

Waldner and Vassileva \cite{waldner_w} trained a different filter per Twitter user, using  Naive Bayes. They classified incoming tweets in three classes (interesting, neutral, uninteresting) and studied user interface designs to emphasize `interesting' tweets in timelines.

\subsection{Hybrid personalized global filters}

Uysal and Croft \cite{uysal_i} consider two tasks: (a) predicting if an incoming tweet will be retweeted by a particular recipient or not and (b) ranking the potential recipients of a particular tweet so that recipients more likely to retweet it will be higher. We considered only the former task, but the same system could be used for the latter task too. The system of Uysal and Croft is hybrid, in the sense that it is global (a single filter for all users), but the feature vectors that represent the tweets include recipient-specific features, as in our own work. The features of Uysal and Croft are also similar to the ones we used. They consider the incoming tweet, the author, the recipient, their previous interaction etc. In fact, our feature set was largely based on that of Uysal and Croft, though we strived for engineering simplicity (e.g., we do not use personal language models), we included additional features (e.g., Klout scores, more similarity scores), and we studied the predictive power (Pearson correlation) of each individual feature, whereas Uysal and Croft assessed the predictive power of entire groups of features only.

Uysal and Croft found that features roughly corresponding to our Group 1 (the tweet itself) were the most useful, whereas in our experiments (Section~\ref{sec:devExperiments}) only Group 2 (content text similarity), Group 3 (influence, authority, popularity), Group 4 (sender-recipient interaction), and Group 6 (neighbours) had features in the top 10. This difference may be due to the different datasets and learning algorithms that we used. Uysal and Croft used a decision tree classifier, whereas we used logistic regression. Also, we used 122 journalists as recipients, whereas Uysal and Croft used 242 random (but reasonably active) Twitter users. On the other hand, the dataset of Uysal and Croft was smaller (24,200 instances in total) compared to ours (133,000 instances), Uysal and Croft did not examine the effect of the size of the training set, and the tweets of their dataset were not temporally ordered. 

Hong et al.\ \cite{Hong2012} use types of features that are similar to the ones we used, but rely on Factorization Machines. We use a much simpler logistic regression classifier, still obtaining very promising results. 

Zhang et al.\ \cite{zhang_q} also developed a hybrid personalized global filter (a single filter for all recipients, with recipient-sensitive feature vectors) to predict retweets. They used word embeddings to represent the words of the tweets and a convolutional neural network (CNN) to construct a single embedding for each tweet. The senders and recipients are also represented by (user) embeddings, and their embeddings influence the behaviour of a second version of the CNN that produces an alternative embedding of each tweet, in effect making the second CNN sensitive to the interests of the senders and recipients. The output tweet embeddings of the two versions of the CNN, concatenated with the embeddings of the recipient and sender and the similarity of the scores of the two CNN versions are then used as a feature vector by a logistic regression classifier layer. The work of Zhang et al.\ is an interesting attempt to avoid manual feature engineering. The embeddings that they use, however, in effect encode information only about the words of the tweet and the previous tweets of the sender and recipient. Our experiments showed that features that consider the influence, authority, and popularity of the sender, the previous interaction between the sender and the recipient, and the neigbours of the recipient are also useful. Their experiments were conducted on a collection of 37,515 incoming tweets from 1,000 random recipients.

\section{Conclusions and future work} \label{sec:conclusions}

We presented a personalized global filter that aims to identify incoming tweets a particular recipient would find interesting enough to retweet. The filter is global in the sense that it is common for all the recipients. It is also personalized in the sense that the incoming tweets are represented as feature vectors that include user-specific features. Thus, the system can produce different predictions per recipient, even for the same incoming tweet, as in personal filters, while still being able to generalize over different users. We experimented with features that examined the content of each tweet, its novelty and its similarity to tweets previously posted or retweeted by the recipient or sender. Furthermore, features describing the network influence and authority of the author and sender, their past interactions and neighbours were used. In experiments with a collection of approximately 130K tweets received by 122 journalists, our system achieved very high accuracy ($F_1 \approx 0.9$) using only 10 features and only 5K training instances. 

Future work could incorporate the features we used (e.g., by turning them into embeddings) in convolutional or recursive neural networks, possibly building upon the work of Zhang et al.\ (\citeyear{zhang_q}). Benchmark datasets are also needed to compare methods proposed by different researchers. The (encoded) dataset of our experiments, which will be made available, is a step towards this direction, but the recipients of its tweets were all journalists.